%% file: quinfo1.tex
\documentstyle[11pt]{article}

\begin{document}
\setlength{\evensidemargin}{-.2cm}
\setlength{\oddsidemargin}{-.2cm}
\setlength{\topmargin}{.5cm}
\newtheorem{theorem}{Theorem}
\newtheorem{definition}[theorem]{Definition}

\begin{center}
{\Large\bf CLASSICAL AND QUANTUM INFO-MANIFOLDS}
\end{center}
\begin{center}
R. F. Streater, Dept. of Maths., King's College London, Strand, WC2R 2LS
\end{center}
\section{Estimation; the Cramer-Rao inequality}

Let $\rho_{\eta}(x)$ be a probability density, depending on a parameter
$\eta\in R$. The {\em Fisher information} of $\rho_\eta$ is defined to be
\cite{Fisher}
\begin{equation}
G:=\int\rho_\eta(x)\left(\frac{\partial \log\rho_\eta(x)}
{\partial\eta}\right)^2 dx.
\end{equation}
We note that this is the variance of the random variable
$Y=\partial\log\rho_\eta/\partial\eta$, which has mean zero. $G$ is associated
with the {\em family} ${\cal M}=\{\rho_\eta\}$ of distributions, rather than
any one of them. This concept arises in the theory of estimation as follows.
Let $X$ be a random variable whose distribution is believed or hoped to
be one of those in ${\cal M}$. We estimate the value of $\eta$ by measuring
$X$ independently $m$ times, getting the data $x_1,\ldots,x_m$. An {\em
estimator} $f$ is a function of $(x_1,\ldots,x_m)$ that is used for this
estimate. So $X$ is a function of $m$ independent copies of $X$, and so is a
random variable. To be useful, the estimator must be independent of $\eta$,
which we do not (yet) know. We say that an estimator is {\em unbiased}
if its mean is the desired parameter; it is usual to take $f$ as a function
of $X$ and to regard $f(x_i)$, $i=1,\ldots,m$ as samples of $f$. Then the
condition that $f$ is unbiased becomes
\begin{equation}
\rho_\eta.f:=\int\rho_\eta(x)f(x)dx=\eta.
\label{unbiased}
\end{equation}
We use the notation $\rho.f$ for the expectation of $f$ in the state $\rho$.
A good estimator should also have only a small chance of being far from the
correct value, which is its mean if it is unbiased.
This chance is measured by the variance. Fisher \cite{Fisher} stated, and
Rao \cite{Rao} and Cramer proved, that the
variance V of an unbiased estimator $f$ obeys the inequality $V\geq G^{-1}$.
For the proof, differentiate eq.~(\ref{unbiased}) w. r. t. $\eta$ to get
\begin{equation}
\int\frac{\partial\rho_\eta(x)}{\partial\eta}f(x)dx=1,
\end{equation}
which can be written as
\begin{equation}
\int Y(x)(f(x)-\eta)\rho_\eta(x)\,dx=\int\left(\frac{\partial
\log\rho}{\partial\eta}\right)\left(f(x)-\eta\right)\rho_\eta(x)\,dx=1.
\end{equation}
We note that this is the correlation
of $Y$ and $f$, so the covariance matrix becomes
\begin{equation}
\left(\begin{array}{cc}
       G&1\\
       1&V
\end{array}\right).
\end{equation}
This is positive semi-definite, giving the result.\hspace{\fill}$\Box$

If we do $N$ independent measurements of the estimator, and average them,
we improve the inequality to $V\geq G^{-1}/N$.
This inequality expresses that, given the family $\rho_\eta$, there is
a limit to the reliability with which we can estimate $\eta$.
Fisher termed $V/G^{-1}$ the {\em efficiency} of the estimator $f$.
Equality in the Schwarz inequality occurs if and only if the two functions
are proportional. Let $-\partial \xi/\partial\eta$ denote the factor of
proportionality. Then the optimal estimator occurs when
\begin{equation}
\log\rho_\eta(x)=-\int\partial\xi/\partial\eta(f(x)-\eta)\,d\eta.
\end{equation}
Doing the integral, and adjusting the integration constant by normalisation,
leads to
\begin{equation}
\rho_\eta(x)=Z^{-1}\exp\{-\xi f(x)\}
\end{equation}
which is the `exponential family'. 

This can be
generalised to any $n$-parameter manifold ${\cal M}=\{\rho_\eta\}$
of distributions, $\eta=(\eta_1,\ldots,\eta_n)$ with $\eta\in R^n$.
Suppose we have unbiased estimators $(f_1,\ldots,f_n)$,
with covariance matrix $V$. Fisher introduced the {\em information
matrix}
\begin{equation}
G^{ij}=\int\rho_\eta(x)\frac{\partial\log\rho_\eta(x)}{\partial\eta_i}
\frac{\partial\log\rho_\eta(x)}{\partial\eta_j}dx.
\end{equation}
We note that $Y^j:=\partial\log\rho/\partial \eta_j$ is a random variable
with zero mean, and that $G^{ij}$ is its covariance matrix.
Rao remarked that $G^{ij}$ provides a Riemannian metric for
${\cal M}$. We now derive the analogue of the inequality when $n>1$.
Put $V_{ij}=\rho_\eta.[(f_i-\eta_i)(f_j-\eta_j)]$, the covariance
matrix of $\{f_i\}$. Differentiate the condition for being unbiased,
\begin{equation}
\int\rho_\eta(x)f_i(x)\,dx=\eta_i
\end{equation}
with respect to $\eta_j$, and rearrange as above, to get
\begin{equation}
\int\rho_\eta(x)Y^i(x)(f_j(x)-\eta_j)\,dx=\delta_{ij}.
\end{equation}
This is the correlation between $Y^i$ and $f_j$. The covariance matrix of
the $2n$ random variables $Y^i,f_j$ therefore is
\begin{equation}
\left(\begin{array}{cc}
      G&I\\
      I&V
\end{array}\right).
\end{equation}
This is therefore a positive semi-definite matrix. If it is not definite, it
has zero as an eigenvalue, which leads to $GV=I$, and the manifold must be
the exponential family, as before. If it is definite,
so is its inverse, which is found to be
\begin{equation}
\left(\begin{array}{cc}
      \left(G-V^{-1}\right)^{-1}&-G^{-1}\left(V-G^{-1}\right)^{-1}\\
      -V^{-1}\left(G-V^{-1}\right)^{-1}&\left(V-G^{-1}\right)^{-1}
\end{array}\right).
\end{equation}
It follows that the leading submatrices $(G-V^{-1})^{-1}$ and
$(V-G^{-1})^{-1}$ are positive definite, and thus so are their inverses.
It follows that we get the matrix inequality $V\geq G^{-1}$.

\section{Entropy methods, exponential families}
Gibbs knew that the state of maximum entropy, given the
mean energy, is the canonical state. More generally, let $\Omega$ be a
countable sample space,
and let $\Sigma$ denote the set of probabilities (or {\em states}) on
$\Omega$. Let
$f_1,\ldots,f_n$ be $n$ linearly independent random variables,
whose means we can
measure. We want to find the `best' choice for the the state,
given these means. The least prejudiced choice of $\rho$ (Jaynes)
is to maximise the entropy $S$ subject to the $n+1$ constraints given by
normalisation and the means of $f_j,\,j=1,\ldots,n$. We use $\lambda,\xi^j$
as Lagrange multipliers; then we must maximise
\[
-\sum_{\omega\in\Omega}\rho(\omega)\log\rho(\omega)-\lambda\sum_\omega
\rho(\omega)-\sum_{j=1}^n\xi^j\rho(\omega)f_j(\omega)    \]
by varying $\rho(\omega)$ subject to no constraints. We get
\begin{equation}
\rho_\xi(\omega)=Z^{-1}\exp-\{\sum_j\xi^jf_j(\omega)\}
\mbox{ where }Z=\sum_\omega\exp\{-\sum_j\xi^jf_j(\omega)\}.
\end{equation}
These make up the {\em exponential manifold {\cal M}} determined by
${\cal F}:={\rm Span}\,\{f_1,\ldots,f_n\}$ and parametrised by
$\xi^1,\ldots,\xi^n$; these are called the canonical coordinates on
${\cal M}$, which has dimension $n$. At least one, say $f_1$,
must be bounded below, to ensure $Z<\infty$ holds for some $\xi$.

The $\xi^j$ are determined by the given expectation values by the conditions
$\rho_\xi.f_j=\eta_j$, $j=1,\ldots,n$.
The $\eta_j$ are thus also coordinates for the manifold (the {\em mixture}
coords.) It is easy to show that
\begin{equation}
\eta_j=-\frac{\partial \Psi}{\partial\xi^j},\hspace{.3in}j=1,\ldots,n;
\hspace{.4in}
V_{jk}=-\frac{\partial \eta_j}{\partial\xi^k},\hspace{.3in}j,k=1,\ldots,m,
\label{metric}
\end{equation}
where $\Psi=\log Z$,
and that $\Psi$ is a convex function of $\xi^j$. The Legendre dual to $\Psi$
is $\Psi-\sum\xi^i\eta_i$ and this is the entropy $S=-\rho.\log\rho$.
The dual relations are
\begin{equation}
\xi^j=\frac{\partial S}{\partial\eta_j}\hspace{.5in}
G^{jk}=-\frac{\partial\xi^j}{\partial \eta_k}.\label{Massieu}
\end{equation}
By the rule for Jacobians, $V$ and $G$ are mutual inverses. Therefore,
the method of maximum entropy leads to the exponential family,
which allows the optimisation of the Cramer-Rao bound, and gives us
estimators of 100\% efficiency.

\section{Manifolds modelled by Orlicz spaces}
Pistone and Sempi \cite{Sempi} have developed a version of information
geometry, which does not depend on a choice of ${\cal F}$, the span of
a finite number of estimators.
Let $(\Omega,\mu)$ be measure space and let ${\cal M}$ be the set of all
probability measures that are equivalent to $\mu$; such a measure is
determined by its Radon-Nikodym derivative $\rho$ relative to $\mu$.
The topology on ${\cal M}$ is not given by the $L^1$-distance, but by an
Orlicz norm.

Given $\rho\in{\cal M}$, the {\em Cramer class} at $\rho$ is the set of
all random variables $X$ on $(\Omega,\mu)$ such that the moment-generating
function
\begin{equation}
\widehat{X}_\rho(t):=\int e^{-tX}\rho d\mu
\end{equation}
is finite in a 'hood of the origin. This is enough to ensure that it
is analytic in an interval about $t=0$. The Cramer class $C_\rho$ at a
point $\rho$ in ${\cal M}$ is furnished with the Luxemburg norm
\begin{equation}
\|X\|_\rho=\inf\left\{r>0:E_\rho\left[\cosh\left(\frac{u}{r}\right)-
1\right]\leq 1\right\}.
\end{equation}
The Cramer class $C$ at $\rho$ is an Orlicz space, and so is a Banach
space with this norm.
The centred Cramer class $C(0)$ is defined as the subset of $C$ at
$\rho$ with zero mean in the state $\rho$; this is a closed subspace.
A sufficiently small ball in the quotient Banach space $C/C(0)$ then
parametrises a 'hood of $\rho$, and can be identified with the tangent
space at $\rho$;
namely, the 'hood contains those points $\sigma$ of ${\cal M}$ such that
\begin{equation}
\sigma=Z^{-1}e^{-X}\rho\hspace{.3in}\mbox{for some }X\in C.
\end{equation}
where $Z$ is a normalising factor.
Pistone and Sempi show that the bilinear form
\begin{equation}
G(X,Y)=E_\rho\left[XY\right]
\end{equation}
is a Riemannian metric on the tangent space $C/C_0$, thus generalising
the Fisher-Rao theory.

This theory is called {\em non-parametric} estimation theory, because we
do not limit the distributions to those specified by a finite number of
parameters, but allow any `shape' for the density $\rho$.
It is this construction that we take over to the quantum case, except that
the spectrum is discrete and the distributions are not always equivalent.

\input quinfo2

%% file: quinfo2.tex
\section{Efron, Dawid and Amari}
A Riemannian metric $G$, eq.~(\ref{Massieu}) gives us
a notion of parallel transport, namely, that given by the Levi-Civita affine
connection. Recall that an affine map, $U$ (acting on the right)
from one vector space ${\cal T}_1$ to another, ${\cal T}_2$, is one that
obeys
\begin{equation}
(\lambda XU+(1-\lambda)YU)=\lambda XU+(1-\lambda)YU,
\mbox{ for all }X,Y\in{\cal T}_1\mbox{ and all }\lambda\in[0,1].
\end{equation}
The same definition works on an {\em affine space}, that is, a convex
subset of a vector space. This leads to the concept of an affine connection. 

Let ${\cal M}$ be a manifold and denote by $T_\rho$ the tangent space at
$\rho\in{\cal M}$. Consider an affine
map $U_\gamma(\rho,\sigma):T_\rho\rightarrow T_\sigma$ defined for each pair of points
$\rho,\sigma$ and each continuous path $\gamma$ in the manifold starting at
$\rho$ and ending at $\sigma$. Let $\rho,\sigma$ and $\tau$
be any three points and $\gamma_1$ a path from $\rho$ to $\sigma$, and
$\gamma_2$ any path from $\sigma$ to $\tau$.
\begin{definition}
We say that $U$ is an {\em affine connection}, if $U_\emptyset=Id$ and
\begin{equation}
U_{\gamma_1\cup\gamma_2}=U_{\gamma_1}\circ U_{\gamma_2}.
\end{equation}
\end{definition}
Let $X$ be a tangent vector at $\rho$; we call $XU_{\gamma_1}$ the parallel
transport of $X$ to $\sigma$, along the path $\gamma_1$.

We also require $U$ to be smooth in $\rho$ in a 'hood of the point
$\rho$, when we identify a ball in the tangent space with part of the
manifold by the exponential map. In physics it is usually the differential
of $U$ along a specified direction that is called `affine connection'.
Equivalently, a connection defines a covariant derivative of a vector
field on the manifold:
\begin{equation}
\nabla_Y X:=d/dt\;XU_\gamma(\rho,\gamma(t))|_{t=0}
\end{equation}
where $\{\gamma(t)\},\;0\leq t\leq 1$ is any path from $\rho$ to $\sigma$,
which starts at $\rho$ in the
direction $Y\in T_\rho$. This is designed to convert vector fields to
tensor fields. Conversely, a covariant derivative defines a connection.
This concept allows us to specify that two tangent vectors to the manifold
at points $\rho$ and $\sigma$ are {\em parallel} if the parallel transport
(along a specified curve) of one from $\rho$ to $\sigma$ is proportional to
the other. A {\em geodesic} is a self-parallel curve on ${\cal M}$: the
tangent vectors to the curve at different points are parallel, when
transported along the curve. Geodesics relative to the Levi-Civita
connection are lines of minimal length, as measured by the metric.

Estimation theory might be considered geometrically as follows. For
theoretical reasons, we expect the distribution of a random variable
to lie on a submanifold ${\cal M}_0\subseteq{\cal M}$ of states. The data
give us a histogram, which is a distribution, but not a pretty one.
We seek the point on ${\cal M}_0$
that is `closest' to the data. Suppose that the sample space is $\Omega$,
with $|\Omega|<\infty$.
Let us place all positive distributions, including the experimental one,
in a common manifold, ${\cal M}$. This manifold will have the Riemannian
structure, $G$, provided by the Fisher metric. We then draw the geodesic
curve through the data point that has shortest distance to the
sub-manifold ${\cal M}_0$; where it cuts ${\cal M}_0$ is our estimate
for the state. This procedure, however, does not always lead to
unbiased estimators. Efron \cite{Efron} and Dawid \cite{Dawid}
noticed that
the Levi-Civita connection is not the only useful one, and that there are
others that might be used in estimation theory. First, the ordinary mixtures
of densities $\rho_1,\rho_2$ leads to
\begin{equation}
\rho=\lambda\rho_1+(1-\lambda)\rho_2,\hspace{.4in}0<\lambda<1.
\label{minus}
\end{equation}
Done locally, this leads to a connection on the manifold, now called
the $(-1)$-Amari connection: two tangents are parallel if they are
proportional as functions on the sample space. This differs from the
parallelism given by the Levi-Civita connection. We need to use
$(-1)$-geodesics to give unbiased estimates for $f$.

There is another obvious convex structure, that obtained from the linear
structure of the space of centred random variables, also known as the
scores. Take $\rho_0\in{\cal M}$ and write $f_0=-\log\rho_0$.
Consider a perturbation $\rho_{_X}$ of $\rho_0$, which
we write as
\begin{equation}
\rho_{_X}=Z_X^{-1}e^{-f_0-X}.
\end{equation}
The random variable $X$ is not uniquely defined by $\rho_X$, since by
adding a constant to $X$, we can adjust the partition function to give the
same $\rho_X$. Among all these equivalent $X$ we can choose the {\em score}
which has zero expectation in the state $\rho_0$: $\rho_0.X=0$.
We can define a sort of mixture of two such perturbed states, $\rho_{_X}$
and $\rho_{_Y}$ by
\begin{equation}
`\lambda\rho_{_X}+(1-\lambda)\rho_{_Y}\mbox{'}:=\rho_{_{\lambda X+
(1-\lambda)Y}}.
\label{plus}
\end{equation}
This is a convex structure on the space of states, and differs from
that given in eq.~(\ref{minus}). It leads to an affine connection, now
called the $(+1)$-Amari connection. How do these connections relate to
the metric?
\begin{definition}
Let $G$ be a Riemannian metric on the manifold ${\cal M}$.
A connection $\gamma\mapsto U_\gamma$ is called a metric connection if
\begin{equation}
G_\sigma(XU_\gamma,YU_\gamma)=G_\rho(X,Y)
\end{equation}
for all tangent vectors $X,Y$ and all paths $\gamma$ from $\rho$ to
$\sigma$.
\end{definition}
The Levi-Civita connection is a metric connection, but the $(\pm)$ Amari
connections are not; they are, however, dual relative to the Rao-Fisher
metric; let $\gamma$ be a path connecting $\rho$ with $\sigma$; then for
all $X,Y$:
\begin{equation}
G_\sigma(XU^+(\rho,\sigma),YU^-(\rho,\sigma))=G_\rho(X,Y).
\end{equation}
Let $\nabla^{\pm}$ be the two covariant derivatives obtained from the
connections $U^{\pm}$. Amari \cite{Amari} defines intermediate covariant
derivatives 
\begin{equation}
\nabla^\alpha=\frac{1}{2}(1+\alpha)\nabla^+ +\frac{1}{2}(1-\alpha)\nabla^-.
\label{Amari}
\end{equation}
These uniquely define connections, $U^{(\alpha)}$, whose dual relative to $G$
is $U^{(-\alpha)}$.
The Levi-Civita covariant derivative is the case $\alpha=0$, which is
self-dual and therefore metric, as is known.
Amari shows that $\nabla^{(\pm)}$ define flat connections without torsion.
Flat means that the transport is independent of the path,
and `no torsion' means that $U$ takes the origin of $T_\rho$ to
the origin of $T_\rho$ around any loop; it is linear, and
not a general affine map. In that case there are affine coordinates,
that is, global coordinates in which the respective convex structure
is obtained by simply mixing coordinates linearly. Amari shows that
for $\alpha\neq\pm1$,
$\nabla^\alpha$ is not flat, but that the manifold is a sphere in
the Banach space $\ell^p$, $p=-\alpha/2+1/2$. In particular, the
case $\alpha=0$ leads to the unit sphere in the Hilbert space $L^2$, and
the Levi-Civita parallel transport is vector translation in this space.
The metric distance between measures is the Hellinger distance, and the
natural coordinates are the square-roots of the densities, imitating the
wave-functions of quantum mechanics.
Similar results were obtained in infinite dimensions in
\cite{Gibilisco,Isola}.

In estimation theory, the method of maximum entropy for unbiased estimators
makes use of the $\nabla^-$ connection. This is true also in the dynamics of
neural nets, dense liquids, Onsager theory,
Brownian particles in a potential and
the Soret and Dufour effects \cite{RFS9};
the micro-state after a small time is replaced by a macrostate, which
is the same as the max-entropy estimation of the state by one on the
manifold generated by exponentials of the macrovariables
(or, slow variables). The (intractible) microdynamics is continuously
projected in a rolling construction onto the (easier) manifold of
exponential states. This idea was proposed by Kossakowski \cite{Kossakowski},
Ingarden, et al. \cite{Ingarden}, and beautifully expounded
by Balian, et al. \cite{Balian}. The resulting non-linear
dynamics can be described thus: after each
time-step of the linear dynamics of the system, Nature makes the best
estimate of the state among those lying on the manifold.

\section{The finite quantum info manifold}
Chentsov \cite{Chentsov} asked whether the Fisher-Rao metric was unique.
Any manifold has a
large number of different metrics on it; apart from those that differ
just by a constant factor, one can multiply a metric by a space-dependent
factor. There are many others. Chentsov therefore imposed conditions on the
metric. He saw the metric (and the Fisher metric in particular) as a measure
of the distinguishability of two states. He argued that if this is to be
true, then the distance between two states must be {\em reduced} by any
stochastic map; for, a stochastic map must `muddy the waters', reducing our
ability to distinguish states. He therefore considered the class of
metrics $G$ that are reduced by any stochastic map on the random variables.
\begin{definition}
A stochastic map is a linear map on the algebra of
random variables that preserves positivity and takes 1 to itself.
\end{definition}
Chentsov was able to prove that the Fisher-Rao metric is unique,
among all metrics, being the only one (up to a constant multiple)
that is reduced by any
stochastic map. It is therefore uniquely defined up to this factor within
the category of commutative function algebras, with stochastic maps as
morphisms.

In quantum mechanics, instead of the abelian algebra of random variables
we use the algebra of matrices $M_n$. Measures on $\Omega$ are replaced by
`states', that is, $n\times n$ density
matrices. For convenience we limit discussion to the interior of the set of
states; these are positive-definite matrices of trace 1, which are faithful
states and invertible matrices. We take this set
to be the manifold ${\cal M}$; it is a genuine manifold, and
not one of the non-commutative manifolds without points that occur in
Connes's theory. The natural morphisms of the quantum info manifold are
the completely positive maps that preserve the identity.
Chentsov found that uniqueness of the metric is not true for quantum
mechanics. (Actually, Petz completed the analysis
after Chentsov died; see \cite{Hasagawa}).

As in the classical case, there are several affine structures on this
manifold. The first comes from the mixing of the states, and is called the
$-1$-affine structure. Coordinates for a state $\rho$ in a hood of $\rho_0$
provided by $\rho-\rho_0$, a small traceless matrix. The whole tangent space
at $\rho$ is thus identified with the set of traceless matrices, and
this is a vector space with the usual rules for adding matrices. Obviously,
the manifold is flat relative to this affine structure.

The $+1$-affine structure is constructed as follows.
Since a state $\rho_0\in{\cal M}$ is faithful we can write
$H_0:=-\log\rho_0$ and any $\rho$ near $\rho_0\in{\cal M}$ as
\begin{equation}
\rho=Z_X^{-1}\exp-(H_0+X)
\end{equation}
for some Hermitian matrix $X$, which is ambiguous up to a multiple of
the identity. We choose to fix $X$ by requiring $\rho_0.X=0$, and call
$X$ the `score' of $\rho$. Then the tangent space at $\rho$ can be
identified with the set of scores, and the $+1$-linear structure is given by
matrix addition of the scores.
Corresponding to these two affine structures,
there are two affine connections, whose covariant derivatives are denoted
$\nabla^{(\pm)}$. Following Hasagawa \cite{Hasagawa2}, one can also
form interpolating affine structures from eq.~(\ref{Amari}).

As an example of a metric on ${\cal M}$, let $\rho\in{\cal M}$, and
for $X,Y$ in $T_\rho$ define the {\em GNS} metric by
\begin{equation}
G_\rho(X,Y)={\rm Re}\,{\rm Tr}[\rho XY].
\end{equation}
This metric is reduced by all cp stochastic maps $F$; that is, it obeys
\begin{equation}
G_{F^*\rho}(XF,XF))\leq G_\rho(X,X),
\end{equation}
in accordance with Chentsov's idea.
$G$ is just the real part of the scalar product in the Gelfand-Naimark-Segal
construction, and is
positive definite since $\rho$ is faithful. This has been adopted by
Helstrom and others \cite{Helstrom,Uhlmann,Ohya} 
in the theory of quantum estimation theory.
However, Nagaoka \cite{Nagaoka} has noted that if we take
this metric, then the $(+1)$ and
the $(-1)$ affine connections are not dual; the dual to the $(-1)$ affine
connection, relative to this metric, is not flat and has torsion. This
failure of duality is confirmed in \cite{Hasagawa}.

In estimation theory we
naturally seek a quantum analogue of the Cramer-Rao inequality.
Given a family ${\cal M}$ of density operators, parametrised by a real
parameter $\eta$, we seek an estimator $X$ whose mean we can
measure in the true state $\rho_\eta$. To be unbiased, we require
${\rm Tr}\,\rho_\eta X=\eta$, which, as in the classical case gives
\begin{equation}
{\rm Tr}\left\{\rho_\eta\rho_\eta^{-1}\frac{\partial\rho_\eta}{\partial\eta}
(X-\eta)\right\}=1.
\end{equation}
It is tempting to regard $L_r=\rho^{-1}\partial\rho/\partial\eta$ as a
quantum
analogue of the Fisher info; it has zero mean, and the above equation says
that its covariance with $X-\eta$ is equal to 1. The Schwarz inequality then
leads to ${\cal V}(X)\geq
[\rho_\eta.(L_r^*L_r)]^{-1}$, where we use $\rho.X$ to denote
${\rm Tr}[\rho X]$. For several estimators, the method 
used earlier gives this as a matrix inequality.

However, $\rho$ and its derivative do not (in
general) commute, so $Y$ is not Hermitian, and is not popular as a measure
of quantum information. Helstrom, and Petz and
Toth \cite{Toth} get round this by using the idea of a {\em logarithmic
derivative}.
Let $g$ be a real or complex scalar product on the space of matrices; we
say that a matrix $L$ is the $g$-logarithmic derivative of the family
$\rho_\eta$ if for any matrix $X$,
\begin{equation}
\frac{\partial\rho_\eta.X}{\partial\eta}=g(L^*,X).
\end{equation}
The symmetric logarithmic derivative uses the real part of the
{\em GNS} metric for $g$, so that
\begin{equation}
\frac{\partial}{\partial\eta}{\rm Tr}(\rho_\eta X)=\frac{1}{2}{\rm Tr}
[\rho_\eta(L_sX+XL_s)].
\end{equation}

Another metric in Chentsov's allowed class is the
Bogoliubov-Kubo-Mori metric; let $X$ and $Y$ have zero mean in the state
$\rho$. Then put
\begin{equation}
g_\rho(X,Y)=\int_0^1{\rm Tr}\left[\rho^\alpha X\rho^{1-\alpha}
Y\right]d\alpha.
\end{equation}
This is one of the family of scalar products found by Petz to obey the
Chentsov property (a similar property was proved in \cite{RFS9},
with detailed balance replacing complete positivity).
The corresponding logarithmic derivative, $L_B$, is defined such that
\begin{equation}
\frac{\partial}{\partial \eta}\rho_\eta.X=\int_0^1\rho_\eta^\lambda L_B
\rho_\eta^{1-\lambda}X\,d\lambda
\end{equation}
and is given explicitly by
\begin{equation}
L_B=\int_0^\infty (\lambda+\rho_\eta)^{-1}\frac{\partial\rho_\eta}
{\partial\eta}(\lambda+\rho_\eta)^{-1}d\lambda.
\end{equation}
Each metric leads to a Cramer-Rao inequality, also in matrix form for
several estimators, and some of these are stronger than others
\cite{Toth,Sudar}.

The $BKM$ metric has other desirable properties, apart from entering in Kubo's
`theory of linear response'.
For the metric $g$, the connections with covariant derivatives
$\nabla^{(\pm\alpha)}$ are dual, and there are affine coordinates
for $\nabla^\alpha$, namely, it is the unit sphere in the (finite-dim.)
Banach space ${\cal C}_p$, the Schatten class with norm
$\|X\|_p=\left({\rm Tr}|X|^p\right)^{1/p}$. The case
$p=1/2$, or $\alpha=0$, leads to the Hilbert space of
Hilbert-Schmidt operators, which has been used in \cite{Brody}.
More, the Massieu function $\log Z$ is the generating function for all
the connected Kubo functions, and in particular, the mean is the first
derivative, and the metric is the second, as in eq.~(\ref{metric}).
The entropy is again the Legendre transform of the Massieu function,
and the reciprocal relations of eq.~(\ref{Massieu}) hold. It follows
that the Cramer-Rao inequality for the $BKM$-metric is achieved exactly
for the exponential family, agreeing with the method of maximum entropy.
\input quinfo3

%% file: quinfo3.tex
\section{Araki's expansionals and the analytic manifold}
Araki \cite{Araki} has considered the case where $\rho$ is a {\em KMS}
state on a $W^*$-algebra. He then perturbed the state by adding
bounded operators to the KMS Hamiltonian; the perturbed
KMS state has a convergent Kubo-Mori perturbation expansion,
which defines an analytic function in the Banach space of bounded
perturbations. We \cite{RFS} try to follow
this for unbounded perturbations.

Let $\Sigma$ be the set of density operators on ${\cal H}$, and let
${\rm int}\,\Sigma$ be its interior, the faithful states. We shall deal only
with systems described by $\rho\in{\rm int}\,\Sigma$; this means that for
a free Schr\"{o}dinger particle, or system of such,
we are limited to systems inside a finite volume of real
space. Then we would expect the entropy to be finite. The following
class of states turns out to be tractable.
Let $p\in(0,1)$ and let ${\cal C}_p$, denote the set of operators
$C$ such that $|C|^p$ is of trace class. This is like the
Schatten class, except that we are in the bad case, $0<p<1$,
for which $C\mapsto ({\rm Tr}[|C|^p])^{1/p}$ is only a quasi-norm.
Let 
\begin{equation}
{\cal C}_{<}=\bigcup_{0<p<1}{\cal C}_p.
\end{equation}

One can show that the entropy
\begin{equation}
S(\rho):=-{\rm Tr}[\rho\log\rho]
\end{equation}
is finite for all states in ${\cal C}_<$. We take the underlying
set of the quantum info manifold to be
\begin{equation}
{\cal M}={\cal C}_<\cap{\rm int}\Sigma.
\end{equation}

We shall cover
${\cal M}$ with balls, each belonging to a Banach space, and shall show
that we have a Banach manifold when ${\cal M}$ is furnished with the
topology induced by the norms; for this, the main problem is to
ensure that various Banach norms are equivalent.

Let $\rho_0\in{\cal M}$ and write $H_0=-\log\rho_0+cI$. We choose $c$ so
that $H_0\geq I$, and we write $R_0=H_0^{-1}$ for the resolvent at $0$.
We define a 'hood of $\rho_0$ to be the set of states of the form
\begin{equation}
\rho_V=Z_V^{-1}\exp-\left(H_0+V\right),
\end{equation}
where $V$ is a sufficiently small $H_0$-bounded form
perturbation of $H_0$. The necessary and sufficient condition to be
Kato-bounded is that
\begin{equation}
\|V\|_0:=\|R_0^{1/2}VR_0^{1/2}\|_\infty<\infty.
\label{norm}
\end{equation}
The set of such $V$ make up a Banach space, ${\cal T}(0)$, with
(\ref{norm}) as norm.
The first result is that $\rho_V\in{\cal M}$ for $V$ inside a small ball
in ${\cal T}(0)$. For the proof, let $a$ be the form-bound of $V$, and let
$q_{_V}$ be the form of $H_0+V$. Then we have for some $b\geq 0$,
\begin{equation}
-bI+(1-a)q_0\leq q_V\leq bI +(1+a)q_0.
\label{inequality}
\end{equation}
Let $L$ be any finite dimensional subspace of ${\rm Dom}\,q_0$, and put
\begin{equation}
\lambda(q,L)=\sup\{q(\psi,\psi):\|\psi\|=1,\;\;\psi\in L\}.
\end{equation}
Then the ordered eigenvalues of $q$ are given by
\begin{equation}
\lambda(q,n)=\inf\{\lambda(q,L):\dim L=n\}.
\end{equation}
From (\ref{inequality}) we have for each $L$,
\begin{equation}
-b+(1-a)\lambda(q_0,n)\leq \lambda(q_V,L).
\end{equation}
Since $\lambda(q_0,n)\rightarrow\infty$ with $n$, the spectrum of
$H_V$ is purely discrete. Thus
\begin{equation}
\exp\beta\left(b-(1-a)\lambda(q_0,n)\right)\geq\exp-\beta\lambda(q_V,n).
\end{equation}
Summing over $n$ gives the traces
\[{\rm Tr}e^{-\beta H_V}\leq e^{\beta(b-(1-a)H_0)}\]
which is of trace class for some $\beta<1$ if $a$ is small enough.

We now consider \cite{Ray} the special case when $V$ is an $H_0$-bounded as an
operator; the condition for this is $\|R_0V\|<\infty.$ Then $V$ is also
form-bounded, since
\begin{equation}
\|R_0^{1/2}VR_0^{1/2}\|_\infty\leq\|R_0V\|_\infty<\infty.
\label{bound}
\end{equation}
In this case we can use the larger norm to provide a topology. This is
not equivalent to the topology we get using the norm (\ref{norm});
we are moving from $\rho_0$ in a direction more regular than the general
direction in the tangent space, and this allows us to furnish this slice
of the manifold with a stronger topology.
The state defined by $V$ is given by
\begin{equation}
\rho_V:=Z_V^{-1}\exp-(H_0+V).
\end{equation}
Thus, $V$ and $V+cI$ give rise to the same state; near $\rho_0$
the regular directions in ${\cal M}$ are thus
parametrised by the quotient space 
\begin{equation}
\widehat{\cal T}={\cal T}/\{cI\}.
\end{equation}
We may therefore use the {\em score},
$V-\rho_0.V$, as coordinates for the `regular' manifold, now using
just the
operator bounded perturbations. We show that these are displacements of the
state in {\em analytic directions}; in \cite{Grasselli}
we find a more general class of analytic directions, which together make up
the `analytic' manifold. This is an attempt to find the quantum analogue
of the Cramer class. We shall come to this later.

The norms $\|R_0V\|_\infty$
on overlapping regions are equivalent. For, around $\rho_V$ we perturb
with $X$ such that $\|R_VX\|_\infty<\infty$, and
\begin{equation}
\|R_VX\|_\infty=\|R_VH_0R_0X\|_\infty\leq\|R_VH_0\|.\|R_0V\|_\infty,
\end{equation}
and the converse inequality holds similarly.
We define the $(+)$-affine connection by transporting the score
$V-{\rm Tr}\,\rho V$ at the point $\rho$ to the score $V-{\rm Tr}\,\sigma V$
at $\sigma$. This connection is flat and torsion-free, since it patently
does not depend on the path between $\rho$ and $\sigma$.
The $(-)$-connection
can be defined in ${\cal M}$ since each ${\cal C}_p$ is a vector space.
It is likely, but not proved, that the $(-)$-mixture of states is continuous
in the topology we have defined here.

A case between operator bounded and form bounded is $\epsilon$-bounded:
\begin{equation}
\|V\|_\epsilon:=\|R_0^{1/2-\epsilon}VR_0^{1/2+\epsilon}\|_\infty<\infty,
0\leq\epsilon\leq 1/2.
\end{equation}
This is the analogue of the Cramer class, since we prove that
$Z$ is an analytic function of $V$ in this case.

Araki proved that if $V$ is bounded, the Kubo-Mori expansion converges:
\begin{equation}
\log Z_V=\sum_{n=0}^\infty(n!)^{-1}\int_0^1\prod d\alpha_i\delta(
\sum\alpha_i-1)K_n
\end{equation}
where
\begin{equation}
K_n:={\rm Tr}\left(\rho^{\alpha_1}V\ldots\rho^{\alpha_n}V\right).
\end{equation}
We prove (with Grasselli) that the series converges also for $\epsilon$-
bounded perturbations, and that the $\|V\|_\epsilon$ are equivalent
on overlapping regions. We now give an outline of the method.

We need an economical estimate for the $n$-Kubo function. If $V$ were
bounded, we could use the H\"{o}lder inequality for traces,
with $p_i=1/\alpha_i$ using that $\sum\alpha_i=1$:
\begin{equation}
|{\rm Tr}\left[\rho^{\alpha_1}V_1\ldots\rho^{\alpha_n}V_n\right]|\leq
{\rm Tr}\,\rho\|V_1\|_\infty\ldots\|V_n\|_\infty.
\end{equation}
We do better, since there is $\beta<1$ such that $\rho^\beta$ is of
trace class, so we can replace $\rho$ by $\rho^\beta$. We can thus borrow
$\rho^{(1-\beta)\alpha_j}$ to help bound the potentials.
Also, as $\sum\alpha_j=1$, the region of integration is the (overlapping)
union of regions $S_j$ where $\alpha_j\geq 1/n$. By cyclicity, we may
take $j=n$. We then write $\rho^{\alpha_j}V_j$ as
\begin{equation}
...\left[\rho^{\alpha_j\beta}\right]\left[H^{1-\delta_{j-1}+\delta_j}
\rho^{(1-\beta)\alpha_j}\right]\left[R^{\delta_j}V_jR^{1-\delta_j}\right]
...
\end{equation}
The dots are factors taken with other terms.
We bound the middle $[...]$ by the spectral theorem, arranging the
parameters $\delta_j$ so that we get an integrable
function of $\alpha_j$ in $S_n$, $1\leq j\leq n-1$. We bound the
final $[...]$ using the $\epsilon$-boundedness of $V$, by a suitable choice
of the $\delta_j$. We end up with a factorial bound on the $n$-point
function, so the series converges as a geometric series.

The manifold can be furnished by a real-analytic structure, by asserting
that the ring of germs of analytic functions on the manifold consists of
functions that are analytic in these analytic directions.
The mixture coordinates $\eta$ are examples of analytic functions; we say
that we have an {\em analytic parametrisation}
of the manifold by $\eta$. It
remains to prove that the $\xi$ are analytic functions of $\eta$, before we
can say that $\eta$ are analytic coordinates.

\section{Singular perturbations}
Every point of our manifold has some directions in its tangent space
that remain within ${\cal M}$ but are not analytic directions.
Consider the anharmonic oscillator, 
\begin{equation}
H=(p^2+q^2)/2+\lambda q^{2n},\hspace{.3in}\lambda>0.
\end{equation}
It is known that $\exp-\beta H$ is of trace-class for all $\beta>0$,
so these states are in ${\cal M}$. It is also known that there is a
singularity at $\lambda=0$. Our result shows that if we start at
$\lambda>0$ then there is a region around this state where the
manifold has analytic directions. Obviously, any point in ${\cal M}$
has many analytic directions: the bounded perturbations,
provide many such. The metric is finite in a much wider class of directions:
if $\rho^\beta$ is of trace-class, and $V$ is a form such
that $\rho^\delta V$ is bounded for $\delta=(1-\beta)/2$, the a
regularised BKM metric in the $V$-direction is finite at $\rho$.

The natural class of states, the analogue of the Orlicz space of
\cite{Sempi}, is the set ${\cal M}_{\rm max}$ of states of finite
entropy. The natural class
of states $\sigma$ in a 'hood of a state $\rho$ of finite entropy consists
of states of finite entropy whose entropy relative to $\rho$ is also finite.
This 'hood will consist of many non-analytic perturbations of $\rho$.
It is known that the $-1$-mixture (the usual mixture) of states of finite
entropy has finite entropy, so ${\cal M}_{\rm max}$ has the $-1$-affine
structure. Here is a simple proof.
\begin{theorem}
\begin{equation}
S(\lambda\rho+(1-\lambda)\sigma)\leq\lambda S(\rho)+(1-\lambda)S(\sigma)
+\lambda\log(1/\lambda)+(1-\lambda)\log(1/(1-\lambda)).
\end{equation}
\end{theorem}
Proof.\\
$-\log x$ is an operator monotone decreasing function. Since $\lambda\rho+
(1-\lambda)\sigma\geq\lambda\rho$, we have
\[ -\log(\lambda\rho+(1-\lambda)\sigma)\leq -\log(\lambda\rho).\]
Hence 
\[-\lambda\rho.\log(\lambda\rho+(1-\lambda)\sigma)\leq-\lambda\rho.\log
(\lambda\rho).\]
Similarly
\[-(1-\lambda)\log(\lambda\rho+(1-\lambda)\sigma)\leq-(1-\lambda)\sigma\log
((1-\lambda)\sigma).\]
Adding, gives
\begin{eqnarray*}
S(\lambda\rho+(1-\lambda)\sigma)&\leq&-\lambda\rho.(\lambda\rho)-
(1-\lambda)\sigma.\log((1-\lambda)\sigma)\\
&=&\lambda S(\rho)+(1-\lambda) S(\sigma)
+\lambda\log(1/\lambda)+(1-\lambda)\log(1/(1-\lambda))<\infty.
\end{eqnarray*}
So the space ${\cal M}_{\rm max}$ of density matrices of finite entropy is
a $(-1)$-affine space.

In \cite{RFS}
we propose a Luxemburg norm for the tangent space at a point
$\rho\in{\cal M}_{\rm max}$. We expect that a 'hood of a point $\rho$ will
consist
of all states $\sigma\in{\cal M}_{\rm max}$ having finite relative entropy,
thus: $S(\sigma|\rho):=\rho.(\log\rho-\log\sigma)<\infty$.\\
\noindent{\bf Acknowledgements}\\
It is a pleasure to thank M. Ohya for the invitation to the conference,
H. Araki for discussions, and H. Hasagawa for arranging the trip.

\end{document}

%% file: quinfo1.bbl
\begin{thebibliography}{99}
\bibitem{Amari}
Amari, S.-I., {\bf Differential Geometric Methods
in Statistics}, {\em Lecture Notes in Statistics}, {\bf 28},
1985. Springer-Verlag.
\bibitem{Araki}
Araki, H., {\em Publ. RIMS}, {\bf 9}, 165-209, Kyoto, 1968.
\bibitem{Balian}
Balian, R., Y. Alhassid and H. Reinhardt, `Dissipation
in many-body systems: a geometrical approach based on information theory',
{\em Phys. Reports}, {\bf 131}, 1-146, 1986.
\bibitem{Brody}
Brody, D. C., and L. P. Hughston, {\em Phys. Lett.}
{\bf 77}, 2851-, 1996.
\bibitem{Chentsov}
Chentsov, N. N., {\bf Statistical Decision and Optimal
Inference}, {\em Nauka}, Moscow, 1972; in Russian. English version,
{\em Amer Math Soc. Translations}, {\bf 53}, 1982.
\bibitem{Dawid}
Dawid, A., `Discussion of a paper by Bradley
Efron', {\em Ann. Stat.}, {\bf 3}, 1231-1234, 1975. `Further comments on a
paper by Bradley Efron', {\em Ann. Stat.}, {\bf 5}, 1249, 1977.
\bibitem{Efron}
Efron, B. `Defining the curvature of a statistical problem',
{\em Ann. Stat.}, {\bf 3}, 1189-1242, 1975. `The geometry of exponential
families', {\em Ann. Stat.}, {\bf 5}, 457-458, 1977.
\bibitem{Fisher}
Fisher, R. A., `Theory of statistical estimation',
{\em Proc. Camb. Phil. Soc.}, {\bf 22}, 700-725, 1925.
\bibitem{Gibilisco}
Gibilisco, P., and G. Pistone, `Connections on
non-parametric statistical manifolds by Orlicz space geometry', {\em
Infinite-dimensional Anal., Quantum Prob., and Related Topics}, {\bf 1},
325-347, 1998.
\bibitem{Isola}
Gibilisco, P., and T. Isola, `Connections on
statistical manifolds of density
operators by geometry of non-commutative $L^p$-spaces, {\em
Infinite-dimensional Analysis, Quantum Probability and Related Topics},
{\bf 2}, 169-178, 1999.
\bibitem{Grasselli}
Grasselli, M., and R. F. Streater, `The quantum
info manifold for epsilon-bounded forms', submitted to {\em Reports on
Math. Phys.}; Los Alamos Archive Math-ph/9910031.
\bibitem{Hasagawa2}
Hasagawa, H. {\em Reps. on Math. Phys}, {\bf 33},
87-, 1993. `Noncommutative extension of the information geometry',
pp 327-337 in {\bf Quantum Communication and Measurement}, eds.
V. P. Belavkin, O. Hirota and R. L. Hudson, Plenum Press, N. Y. 1995.
\bibitem{Hasagawa}
Hasagawa, H., and D. Petz, `Non-commutative extension of
information geometry II', 109-118 in {\bf Quantum Communication, Computing
and Measurement}, Eds. O. Hirota et al., Plenum Press, N. Y. 1997.
\bibitem{Helstrom}
Helstrom, C. W., {\bf Quantum Detection and Estimation
Theory}, Academic Press, N. Y., 1976.
\bibitem{Ingarden}
Ingarden, R., Y. Sato, K. Sagura, and
T. Kawaguchi, `Information thermodynamics and differential geometry'
{\em Tensor}, {\bf 33}, 347-353, 1979.
\bibitem{Kossakowski}
Kossakowski, A., `On the quantum informational
thermodynamics', {\em Bull. acad. polonaise des sciences}, {\bf 17},
263-267, 1969.
\bibitem{Nagaoka}
Nagaoka, H., `Differential aspects of quantum state
estimation and relative entropy', in {\bf Quantum Communication and
Measurement}, eds. V. P. Belavkin et al., Plenum Press, 1995.
\bibitem{Ohya}
Ohya, M. and D. Petz, {\bf Quantum Entropy and its Use},
Springer-Verlag, 1993.
\bibitem{Toth}
Petz, D., and G. Toth, `The Bogoliubov inner product in
quantum statistics', {\em Lett. in Math. Phys.}, {\bf 27}, 205-216, 1993.
\bibitem{Sudar}
Petz, D., and C. Sudar, `Geometries of quantum states',
{\em J. Mathematical Phys.}, {\bf 37}, 2662-2673, 1996.
\bibitem{Sempi}
Pistone, G., and C. Sempi, `Infinite-dimensional
geometric structure on the space of all probability measures
equivalent to a given one', {\em Annals of Statistics}, {\bf 33},
1543-1561, 1995.
\bibitem{Rao}
Rao, C. R., `Information and accuracy attainable in the
estimation of statistical parameters', {\em Bull. Calcutta
Math. Soc.}, {\bf 37}, 81-91, 1945.
\bibitem{RFS9}
Streater, R. F., `Gas of Brownian particles in a potential', {\em J. Stat.
Phys.}, {\bf 88}, 447-, 1997.
`Information geometry and reduced
quantum description', {\em Reports on Math. Phys.}, {\bf 38}, 419-436, 1996.
`A model of dense liquids', {\em Banach Center Publications},
{\bf 43}, 381-393, Warsaw, 1998. `Onsager relations in statistical dynamics'
{\em Open Systems and Info. Dyn}, {\bf 6}, 87-100, 1999. `The Soret and
Dufour effects in statistical dynamics', {\em Proc. Roy. Soc.}, {\bf 456},
205-221, 1999. Los Alamos Archive math-ph/9910043
\bibitem{RFS}
Streater, R. F. `The information manifold for relatively
bounded potentials', to appear in the Bogoliubov Memorial Volume, ed.
A. A. Slavnov, Steklov Institute, Moscow; 2000. Los Alamos Archive Math-ph
9910035.
\bibitem{Ray}
Streater, R. F.,`The analytic quantum info manifold',
to appear in {\bf Stochastic Processes, Physics and Geometry}, eds.
F. Gesztesy, S. Paycha and H. Holden; Canad. Math, Soc., 2000.
Los Alamos Archive Math-ph/9910036.
\bibitem{Uhlmann}
Uhlmann, A., `The metric of Bures and the geometric phase'
267-274, in {\bf Groups and Related Topics}, eds. R. Gielerak et al., Kluwer,
1992. `Density operators as an arena for differential geometry', {\em Reps.
Math. Phys.}, {\bf 33}, 253-263, 1993.

\end{thebibliography}
